# NEW INVESTIGATION OF THE EVAPORATION OF MULTICOMPONENT HYDROCARBONS LIQUID DROPLETS IN ROTATORY FORCED CONVECTION

J. Dgheim[*†], R. al Maarrawi[†], M. Abdallah[†], N. Nasr[†]

† Laboratory of Applied Physics (LPA), Group of Mechanical, Thermal & Renewable Energies (GMTER)

**Abstract**: Heat and mass transfers of the evaporation of rotating hydrocarbons liquid droplet of ternary components, in forced convection are studied numerically. Comparison between our numerical results and those obtained in the literature shows qualitative and quantitative satisfactory agreements, in rotatory forced convection. Optimal values of the wind and rotational velocities of the liquid droplet have been determined. New correlations expressing the evaporation rate in terms of Dgheim & Reynolds numbers are proposed. Other new correlations expressing the evaporation rate in terms of Dgheim, Reynolds, Prandtl and Schmidt numbers are realized. These correlations take into consideration the evaporation of the liquid droplet of ternary components in variable ambient medium temperatures, variable mass fractions, variable air velocities, and variable droplet initial radii, in rotatory forced convection.

[*] Author to whom correspondence should be addressed, e-mail : jdgheim@ul.edu.lb, Fax : +961.1.68.15.53

## 1. Introduction

The phenomena of the fuel droplet evaporation have been studied by many researchers since several years. The evaporation process within an engine is normally accomplished by the injection of the fuel as a spray of liquid droplets. However, its technical applicability is in general complex, due to the interactions between its various physical mechanisms. Thus, the evaporation of a single droplet is frequently considered, as reported by (Abramzon and Sirignano, 1987) and (Dgheim et al, 2013a). Consequently, it is difficult to isolate and study individual physical process, mainly in environments where complicating factors exist such as the spray evaporation phenomenon. Hence, it is useful to study the simplest configurations of the evaporation phenomenon. The numerical models of the single droplet evaporation have been considered in the last twenty years using more developed numerical and experimental techniques.

In general, the evaporation and the combustion phenomena of the liquid fuels are blends of several chemical products. Each of these products is characterized by its volatility, diffusivity and thermal conductivity. These thermo-physical properties of the multi-component fuel are used in many industrial systems such as the production of new chemical products. The simplest multi-component fuels are, in general, binary or ternary. It exhibits phenomena that are not visible with the pure fuels. These fuels possess characteristics of multi-component blends. The difference in volatility and the species diffusion in liquid phase are essential for the evaporation and combustion in multi-component blends. The evaporation phenomena of an isolated droplet of multi-component fuel have been studied by (Dgheim et al, 2005b and 2012c) and (Sazhin, 2006a).

The regression of the square radius and the evolution of the surface temperature are considered by several researchers such as (Continillo and Sirignano, 1998), (Cho and Dryer, 1999), (Bouaziz et al, 2001), and (Merouane and Bounif, 2010). These authors studied numerically and experimentally the evaporation of hydrocarbon droplets of binary components in natural and forced convections. The authors verified the $d^2$ law. On the other hand, their model depends only on the initial composition of the fuel droplets. (Aharon and Shaw, 1998) considered experimentally the evaporation and the combustion of heptane and hexadecane mixture droplets in a reduced gravity environment. The author’s model shows that the liquid species diffusion coefficient varies significantly. This is due to the variations in droplet temperatures. (Sazhin et al, 2010b) developed a simplified model concerning the bi-component droplet heating and evaporation. Sazhin’s model applies the mean droplet temperatures in a mono-disperse spray, in particular, for the pure acetone and acetone-rich mixture droplets. Moreover, their simulation shows good agreement of the temperatures distribution predicted by the simplified model with the earlier reported model. In addition, the temperatures distributions predicted by the ideal model do not diverge more than few degrees from the non-ideal model.

The droplet's burning rate of a mono-component liquid is examined by (Xu et al, 2003a and 2004b). They proposed a correlation for the droplet's burning rate as function of Reynolds number. They studied the influence of the room temperature, and the droplet initial radius on the mentioned rate. However, the influence of the droplet evaporation rate on the evaporation phenomenon near the rotating droplet has not been studied yet.

The topic of this research, which is the following work on the evaporation rate of the rotating liquid droplets ((Dgheim et al, 2017d, 2018e and 2018f) has not been deeply developed by the researchers, despite their multiple applications and useful for the scientific committee. However, this field has not been the subject of intense study due to its complicated mathematical formulations.

## 2. Mathematical model

Our mathematical model used the heat and mass transfers' equations in the liquid phase of the hydrocarbon liquid droplet. The isolated droplet is considered as a saturated sphere in rotation with binary and ternary components. It is evaporated in a hot medium under the effect of the rotatory forced convection. The mathematical equations are solved numerically using an implicit finite difference method. The numerical method is used to scheme the mathematical equations describing the evaporation of a rotating droplet of three components, namely, heptane, octane and decane.

Figure 1 shows a liquid fuel droplet of radius $r_s$. The rotating droplet is exposed to a hot air in forced convection, where ($O$) is the origin of the droplet center, ($x$) is the curvilinear direction, and ($y$) is the ordinate considered positive towards the outside of the rotating liquid droplet. This figure considers the model proposed by Abramzon and Sirignano. Nevertheless, the model is extended to the vaporization of the liquid droplets of binary and ternary components of the rotating liquid droplet. Further, the model is based on the film theory. In this theory, heat and mass transfers between the droplet surface and the external gas happens inside a thin gaseous film surrounding the rotating droplet. Moreover, it assumes laminar regime and do not consider surface tension, chemical reaction, radiation, Soret and Dufour effects.

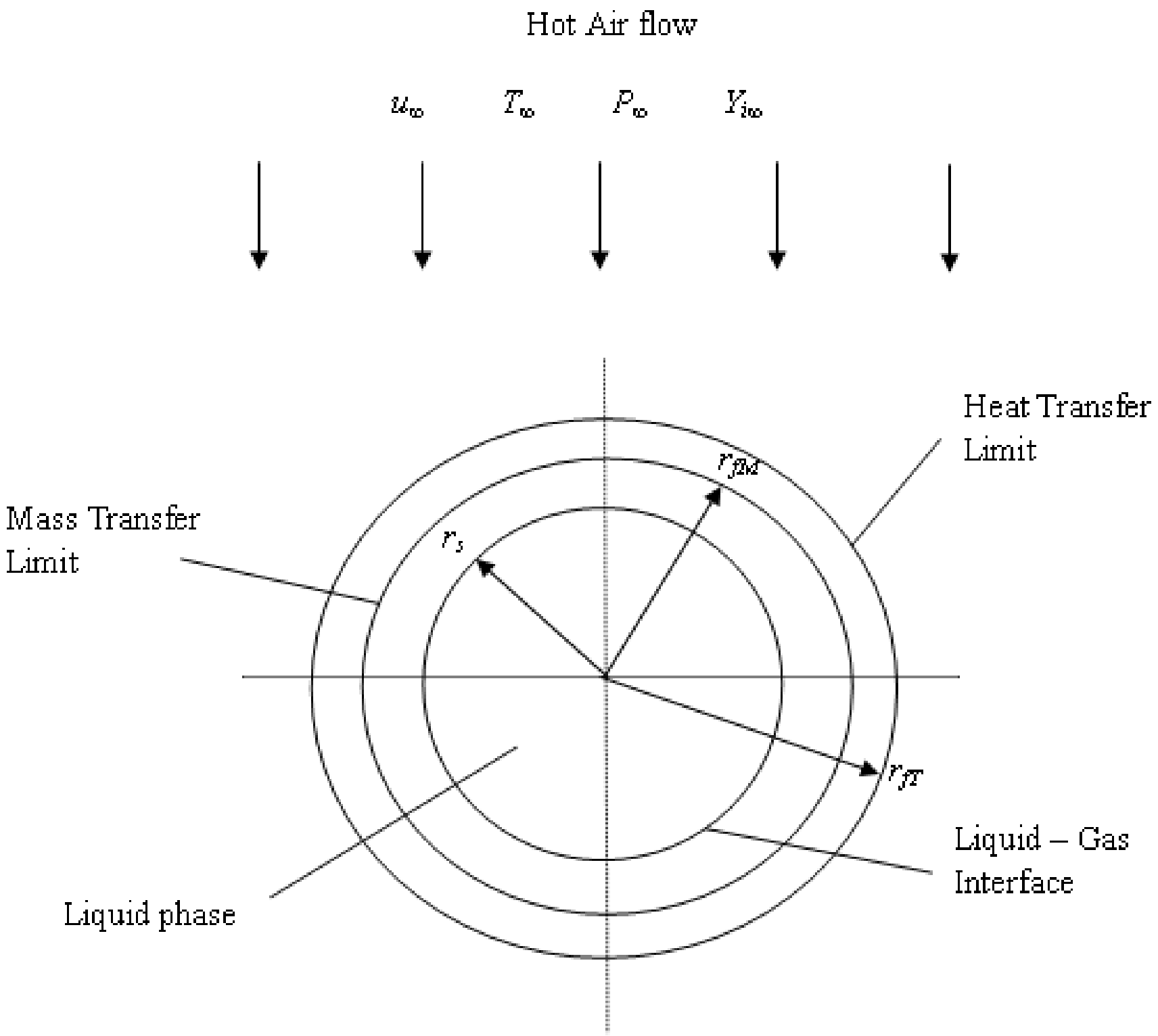

Figure 1: The physical model

The mathematical model is used to determine the distribution of the droplet surface temperature and the regression of the square radius. The thickness of the gaseous film depends on the average Nusselt and Sherwood numbers. The film thickness limits the integration areas of the mass and energy conservation equations in the gaseous phase.

Energy equation:

$$\rho Cp\frac{\partial T_L}{\partial t} = \frac{1}{r^2}\frac{\partial}{\partial r}\left(\lambda r^2 \frac{\partial T_L}{\partial r}\right) \tag{1}$$

Where ($\rho$) is the density, ($Cp$) is the specific heat, ($T_L$) is the liquid hydrocarbon temperature, ($t$) is the time, ($r$) is the radial direction, and ($\lambda$) is the thermal conductivity.

Diffusion equation:

$$\rho\frac{\partial Y_{L,i}}{\partial t} = \frac{1}{r^2}\frac{\partial}{\partial r}\left(r^2 D \frac{\partial Y_{L,i}}{\partial r}\right) \tag{2}$$

Where ($Y_{L,i}$) is the liquid mass fraction of the species ($i$), and ($D$) is the diffusion coefficient.

The energy equation is used to compute the rotating droplet surface temperature versus time. The mass diffusion equation is used to determine the radius regression, which is computed using the evaporated mass debit of the rotating liquid droplet.

Initial conditions: $\forall\ t < t_0$, ($t_0$) corresponds to the beginning of the evaporation process.

At the liquid droplet surface, the temperature and mass fraction are supposed to be equal respectively to the rotating droplet surface temperature ($T_s$) and surface mass fraction ($Y_{s,i}$). The surface mass fraction varies according to the saturated vapor pressure ($P_{vs}$) determined by Clausius Clapeyron equation, as reported by Abramzon and Sirignano (1987). Therefore:

$$T_s = T_{s0}$$

$$Y_{s,i} = Y_{s0,i} = \frac{P_{VS}}{\left( P_{VS} + \frac{(P_\infty - P_{VS}) M_a}{M_f} \right)} \quad (3)$$

Where ($T_{s0}$) and ($Y_{s0,i}$) are the initial surface temperature and initial surface mass fraction of the rotating liquid droplet, ($P_\infty$) is the atmospheric pressure, ($M_a$) and ($M_f$) are the molar mass of the air and the fuel respectively.

Inside the liquid phase:

$$T_L = T_{L0}$$

$$Y_{L,i} = Y_{L0,i}$$

Where ($T_{L0}$) and ($Y_{L0,i}$) are the initial liquid temperature and initial liquid mass fraction of the rotating liquid droplet respectively.

The latent heat of vaporization is a function of the surface temperature variation, given by:

$$L = L_v \left( \frac{(T_{crit} - T_s)}{(T_{crit} - T_{ebn})} \right)^{0.38} \quad (4)$$

Where ($L_v$) represents the fuel latent heat of vaporization as reported by (Dgheim and Zeghmati, 2005f), ($T_{crit}$) and ($T_{ebn}$) are the critical and boiling temperatures respectively.

Boundary conditions: $\forall\ t > t_0$

At the droplet center ($r$ = 0):

$$\left. \frac{\partial T_L}{\partial r} \right|_{r=0} = 0$$

$$\left.\frac{\partial Y_{L,i}}{\partial r}\right|_{r=0} = 0$$

At the droplet surface ($r = r_s$):

$$\lambda_g \left.\frac{\partial T_g}{\partial r}\right|_s = \lambda_l \left.\frac{\partial T_L}{\partial r}\right|_s - \rho D L \left.\frac{\partial Y_{L,i}}{\partial r}\right|_s \tag{5}$$

Where ($\lambda_g$) and ($\lambda_l$) are the gas and liquid thermal conductivities respectively, and ($T_g$) is the gas temperature.

The mass fraction at the liquid-vapor interface is a function of the saturated vapor pressure, which varies according to the liquid surface temperature.

In order to avoid the non-uniformity of the mesh spacing near the rotating droplet surface, the variables ($t$, $r$) are transformed to $\left(t,\ \eta = \frac{r}{r_S}\right)$, namely:

$$\left.\frac{\partial}{\partial t}\right|_{t,r} = \frac{\partial}{\partial t} - \frac{\eta}{r_S}\frac{dr_S}{dt}\frac{\partial}{\partial \eta}; \qquad \left.\frac{\partial}{\partial r}\right|_{t,r} = \frac{1}{r_S}\frac{\partial}{\partial \eta}; \qquad \left.\frac{\partial^2}{\partial r^2}\right|_{t,r} = \frac{1}{r_S^2}\frac{\partial^2}{\partial \eta^2} \tag{6}$$

Where ($r_s$), is the surface radius of the rotating liquid droplet.

Therefore, the transfer equations become:

$$\frac{\partial T_L}{\partial t} = \left(\frac{2\lambda_L}{\rho C p \eta r_s^2} + \frac{\eta}{r_s}\frac{dr_s}{dt}\right)\frac{\partial T_L}{\partial \eta} + \frac{\lambda_l}{\rho C p r_s^2}\frac{\partial^2 T_L}{\partial \eta^2} \tag{7}$$

$$\frac{\partial Y_{L,i}}{\partial t} = \left(\frac{2D}{\rho \eta r_s^2} + \frac{\eta}{r_s}\frac{dr_s}{dt}\right)\frac{\partial Y_{L,i}}{\partial \eta} + \frac{D}{\rho r_s^2}\frac{\partial^2 Y_{L,i}}{\partial \eta^2} \tag{8}$$

The boundary conditions become: $\forall\ t > t_0$

At the droplet center ($\eta = 0$):

$$\frac{\partial T_L}{\partial \eta} = 0$$

$$\frac{\partial Y_{L,i}}{\partial \eta} = 0$$

At the droplet surface ($\eta = 1$):

$$\lambda_g \left.\frac{\partial T_g}{\partial \eta}\right|_s = \lambda_L \left.\frac{\partial T_L}{\partial \eta}\right|_s - \rho D L \left.\frac{\partial Y_{L,i}}{\partial \eta}\right|_s \tag{9}$$

In our model, a spinning frequency is introduced to the evaporation of the liquid droplets in forced convection. Thus, Nusselt and Sherwood numbers are determined to be a function of three terms: a first term is related to the evaporation of a stagnant spherical droplet equal to 2, a second term is related to the evaporation of the liquid droplet in forced convection as proposed by (Renksizbulut and Yuen, 1983), and a third term is related to the evaporation of the liquid droplet under the effect of rotation as proposed by (Dgheim et al, 2013g) :

$$Nu^* = 2 + 0.57Re^{1/2}Pr^{1/3}(1+B_T)^{-0.7} + 2.11DG\, Pr^{1/2}(1+B_T)^{0.69} \quad (10)$$

$$Sh_j^* = 2 + 0.87Re^{1/2}Sc_j^{1/3}(1+B_M)^{-0.7} + 1.79DG\, Sc_j^{1/2}(1+B_M)^{-2} \quad (11)$$

Where ($Pr$) is the Prandtl number, ($Sc_j$) is the Schmidt number of the species ($j$), ($B_T$) and ($B_M$) are thermal and mass Spalding numbers respectively, ($Nu^*$) and ($Sh_j^*$) are the modified average Nusselt and Sherwood numbers respectively.

The Dgheim ($DG$) and Reynolds ($Re$) dimensionless numbers are presented as the following:

$$DG = r_s(f/\nu)^{0.5} \quad (12)$$

$$Re = 2ur_s/\nu \quad (13)$$

With ($f$) is the ordinary spinning frequency, ($u$) is the air velocity and ($\nu$) is the kinematic viscosity. Mass and thermal balances are always verified at the droplet's surface. The external radius that limits the integration areas of the conservation equation in the gaseous phase is defined by Abramzon and Sirignano. It depends on the modified Nusselt and Sherwood numbers:

$$r_{fT} = \frac{r_s Nu^*}{Nu^* - 2} \quad (14)$$

$$r_{fM_j} = \frac{r_s Sh_j^*}{Sh_j^* - 2} \quad (15)$$

This model uses the relationships proposed by Abramzon and Sirignano, as the following:

$$Nu = Nu^* \frac{\ln(1+B_T)}{B_T} \quad (16)$$

$$Sh_j = Sh_j^* \frac{\ln(1+B_M)}{B_M} \quad (17)$$

The model allows us to determine the surface temperature and the regression of the square radius of the rotating liquid droplet. The mathematical model is solved by taking into account the variability of the thermo-physical and transport properties in the liquid and vapor phases. The density, viscosity, thermal conductivity, heat capacity, binary diffusion coefficient of the air-fuel vapor mixture, and the latent heat of vaporization are computed using (Heat Atlas Handbook, 1993) and Abramzon and Sirignano. Their works confirm that the thermo-physical properties depend on the temperature and the compositions of the rotating liquid droplet. The simulation computes the thermo-physical and transport properties at the local temperature and local fuel mass fraction, namely:

$$\overline{T}_L = \frac{\left(T_L + T_s\right)}{2} \qquad \overline{Y}_{js} = \frac{\left(Y_j + Y_{js}\right)}{2} \tag{18}$$

## 3. Results and discussion

The numerical model is performed for a range of fuel mixtures, namely, decane, octane and heptane for variable thermo-physical and transport properties. The initial radius of the rotating liquid droplet varies between 0.1 mm and 0.7 mm. The ambient pressure is one atmosphere, whereas the ambient temperature varies between 348 K and 550 K. The wind velocity varies between 0 m/s and 3.1 m/s. The rotational velocity of the liquid droplet remains below 30 rps.

### 3.1. Model accuracy

The numerical results of the evaporation in forced convection, of ternary hydrocarbon droplets (75% heptane - 1% octane - 24% decane) are compared to Bouaziz et al experimental results and to our numerical results of binary components (25% heptane - 75% decane) by fixing the rotational velocity of the liquid droplet to a very small value ($\omega$ =0.00001 rps). After reducing the effect of the rotation phenomenon, the regression of the droplet square radius and the evolution of the droplet surface temperature (*figure 2*) of the whole results show good qualitative and quantitative agreements. Furthermore, good agreements are observed between both binary and ternary components of hydrocarbons liquid droplet, when the initial mass fraction of the octane component is taken equal to 1%. The comparison was performed in a hot ambient medium (348 K), for initial droplet radius of 0.54 mm, and for a wind velocity equal to 3.1 m/s. The relative error is smaller than 10%.

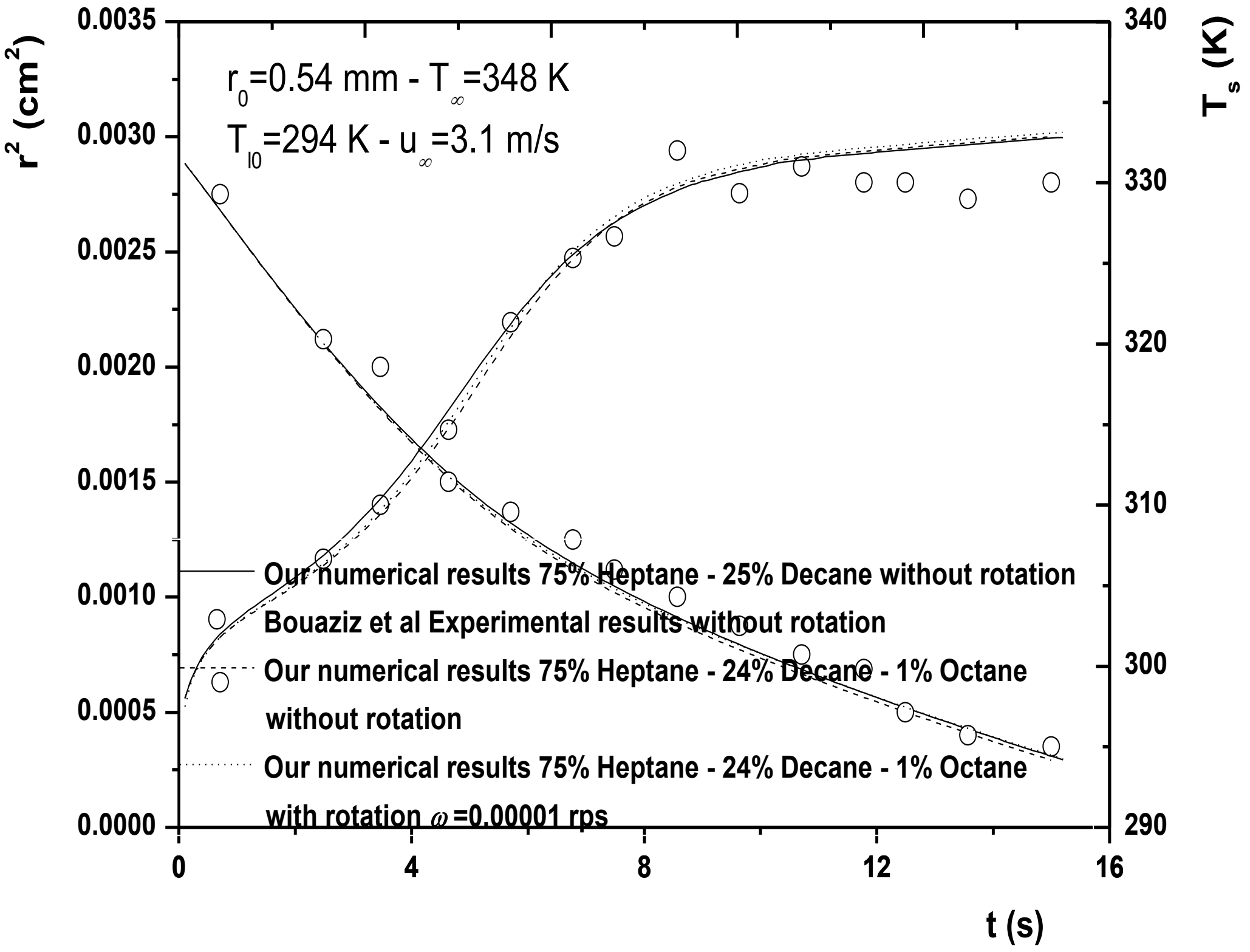

Figure 2: Comparison between our results concerning the droplet square radius regression and the droplet surface temperature evolution of binary and ternary components and the experimental results of Bouaziz et al. Initial radius $r_0$ = 0.54 mm.

In figure 2, one can observe clearly the influence of the volatility of each component on the regression of the square radius of hydrocarbons liquid droplet of binary and ternary components. The stagnant hydrocarbon liquid droplet is evaporated in forced convection. Figure 2 shows also a steep linear decrease of the droplet square radius at the beginning of the evaporation phenomenon. This phenomenon is due to the fast evaporation of the 75% of heptane initial mass fraction, the more volatile component. Then, a further decrease in a slower linear slope is shown due to the evaporation of the 24% of the less volatile component (the decane). Moreover, figure 2 indicates as well an increase of the surface temperature versus time. The surface temperature reaches the saturation temperature of the less volatile component. The droplet surface temperature increases rapidly to reach the heptane boiling temperature. It follows respectively by the evaporation of the octane and decane mass fractions. Thus, this phenomenon shows the non-miscibility of the mixture of the liquid hydrocarbons.

### 3.2. Physical parameters development of the evaporation of rotating hydrocarbon droplet of ternary components

The influence of the physical parameters, such as the droplet initial radius, the mass fraction of the components of the liquid droplet, the wind velocity in forced convection phenomenon, the rotational velocity of the liquid droplet, and the temperature of the surrounding medium, on the evaporation of the rotating hydrocarbon liquid droplet of three components, was studied numerically.

The influence of the rotation of the liquid droplet on the evaporation phenomenon in forced convection, for different initial radii, is presented in figure 3. When the initial radius increases, the evaporation time increases as well. Nevertheless, the evaporation time is remarkably reduced by taking into consideration the rotation of the liquid droplet. Thus, the rotation of the liquid hydrocarbon droplet of ternary components accelerates and improves the evaporation process.

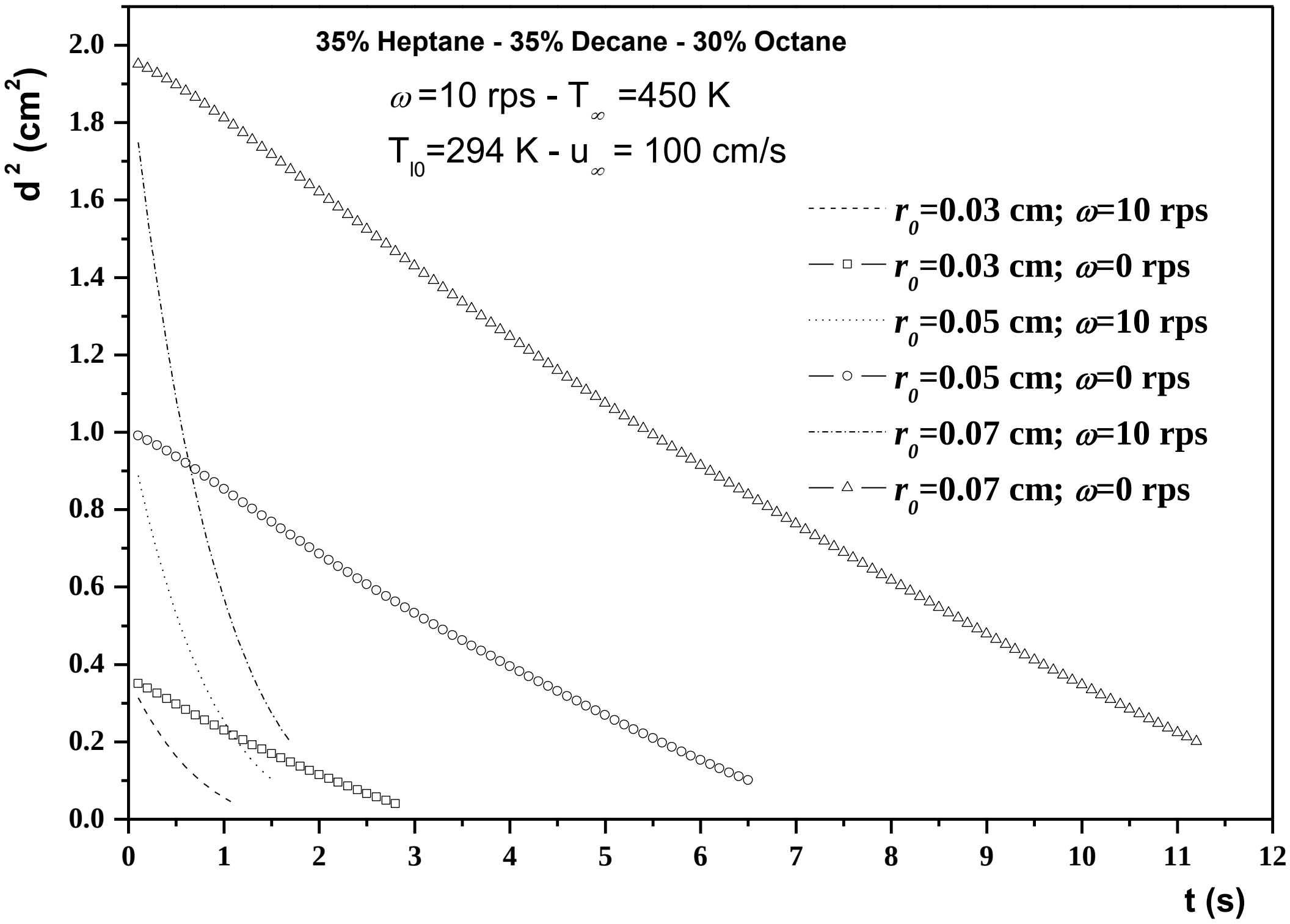

Figure 3: Comparison between our results concerning the droplet square diameter regression and without rotation.

Figures 4 and 5 show the influence of the rotational velocity of the liquid droplet of three components, on the evaporation phenomenon, in forced convection. These figures show that the increasing of the rotational velocity of the liquid droplet accelerates the evaporation phenomenon by reducing the time of the evaporation process. This phenomenon is explained by a quick rise of the droplet surface temperature (*figure 4*). The values of the surface temperature remain below the boiling temperature of the less volatile component. In addition, Dgheim dimensionless number increases with the increasing of the rotational velocity (*DG*=0.223 for $\omega$=30 rps).

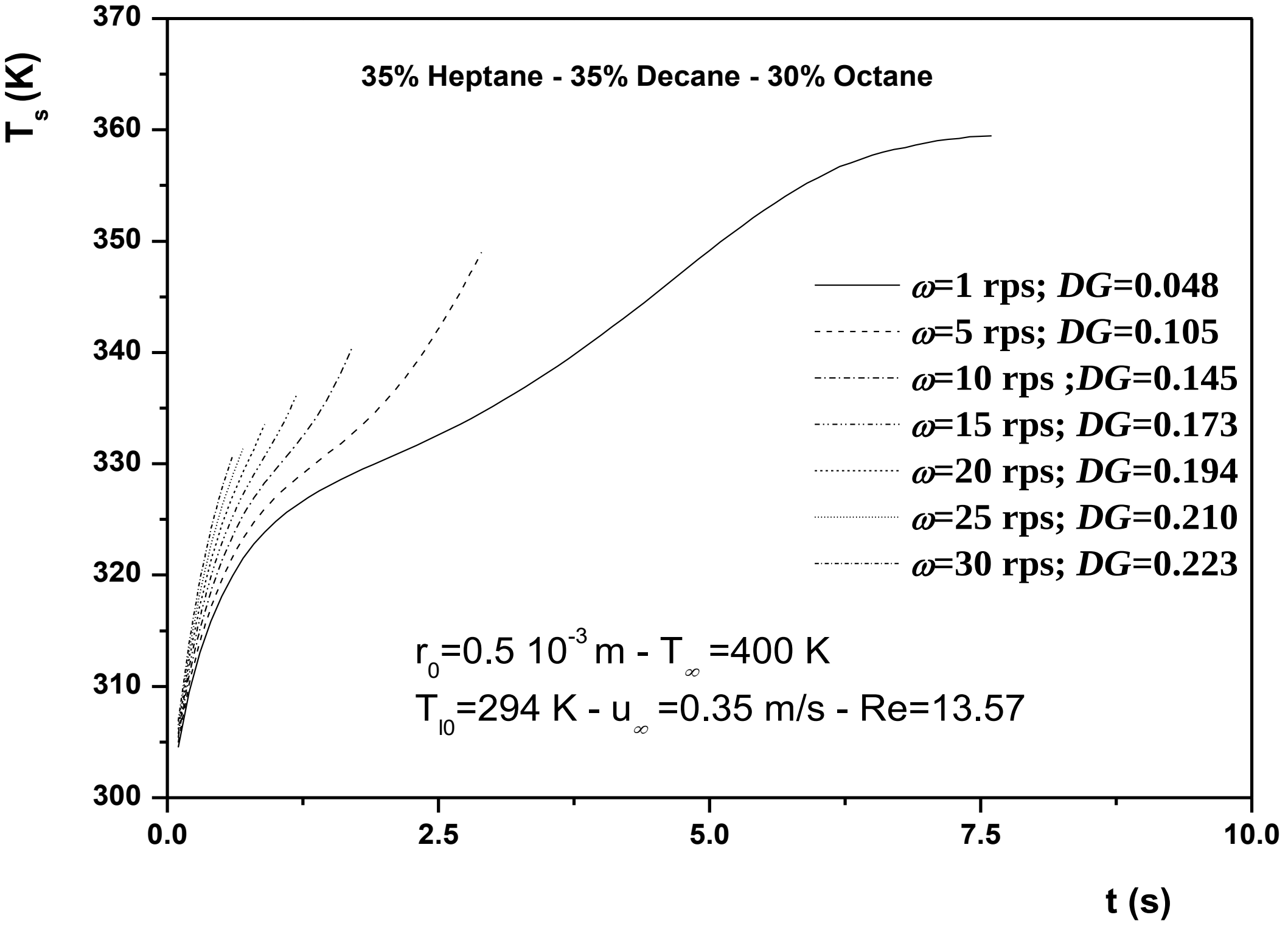

Figure 4: Time increase of the droplet surface temperature of ternary components (35% Heptane - 35% Decane – 30% Octane) in evaporation in forced convection for different rotating velocities.

Therefore, the figure 5 shows that the rise of the rotational velocity evaporates rapidly the multi-component liquid droplet. Furthermore, when the rotational velocity increases from 1 to 5 rps, the slope of the curve becomes less inclined. The inclination of the curve's slope stabilizes when the rotational velocity becomes greater than 15 rps. Thus, the optimal value of the rotational velocity of the three components hydrocarbon droplet is equal to 15 rps. Finally, the optimal value of the rotational velocity accelerates and improves the evaporation process.

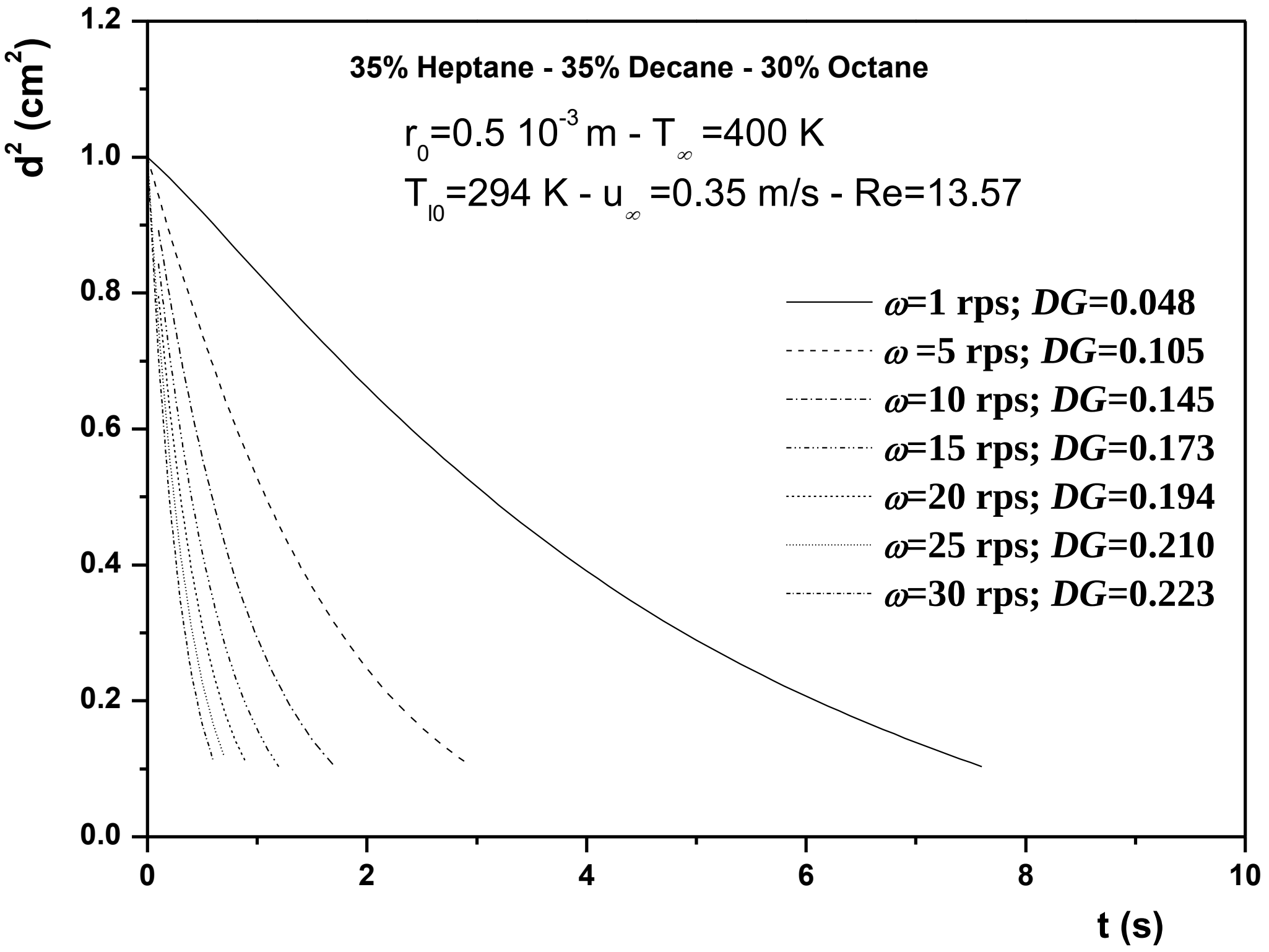

Figure 5: Time increase of the regression of the droplet square diameter of ternary components (35% Heptane - 35% Decane - 30% Octane) in evaporation in forced convection for different rotating velocities.

The influence of the ambient temperature on the evaporation of the rotating hydrocarbon droplet of ternary components, in forced convection, is shown in figures 6 and 7, for low wind velocity. The Reynolds number, characterizing the flow regime near the liquid droplet, is determined using the initial diameter of the liquid droplet, and the thermo-physical properties of the wind. Therefore, the low values of the Reynolds number decreases with the augmentation of the ambient temperature. In addition, the increasing of the ambient temperature changes significantly the surface temperature and the square radius profiles of the rotating hydrocarbons liquid droplet. Figure 6 shows the three stages of the profile of the surface temperature of the rotating hydrocarbon droplet. The first stage is formed by a fast increase of the surface temperature to reach the boiling temperature of the most volatile component. The second stage consists of a moderate increase of the surface temperature to reach the boiling temperature of the octane component. Finally, a third stage is completed by reaching the surface temperature of the less volatile component, the decane. The rise of the surface temperature of the liquid droplet is due to the high diffusivity of each component into the other that depends on the ambient temperature.

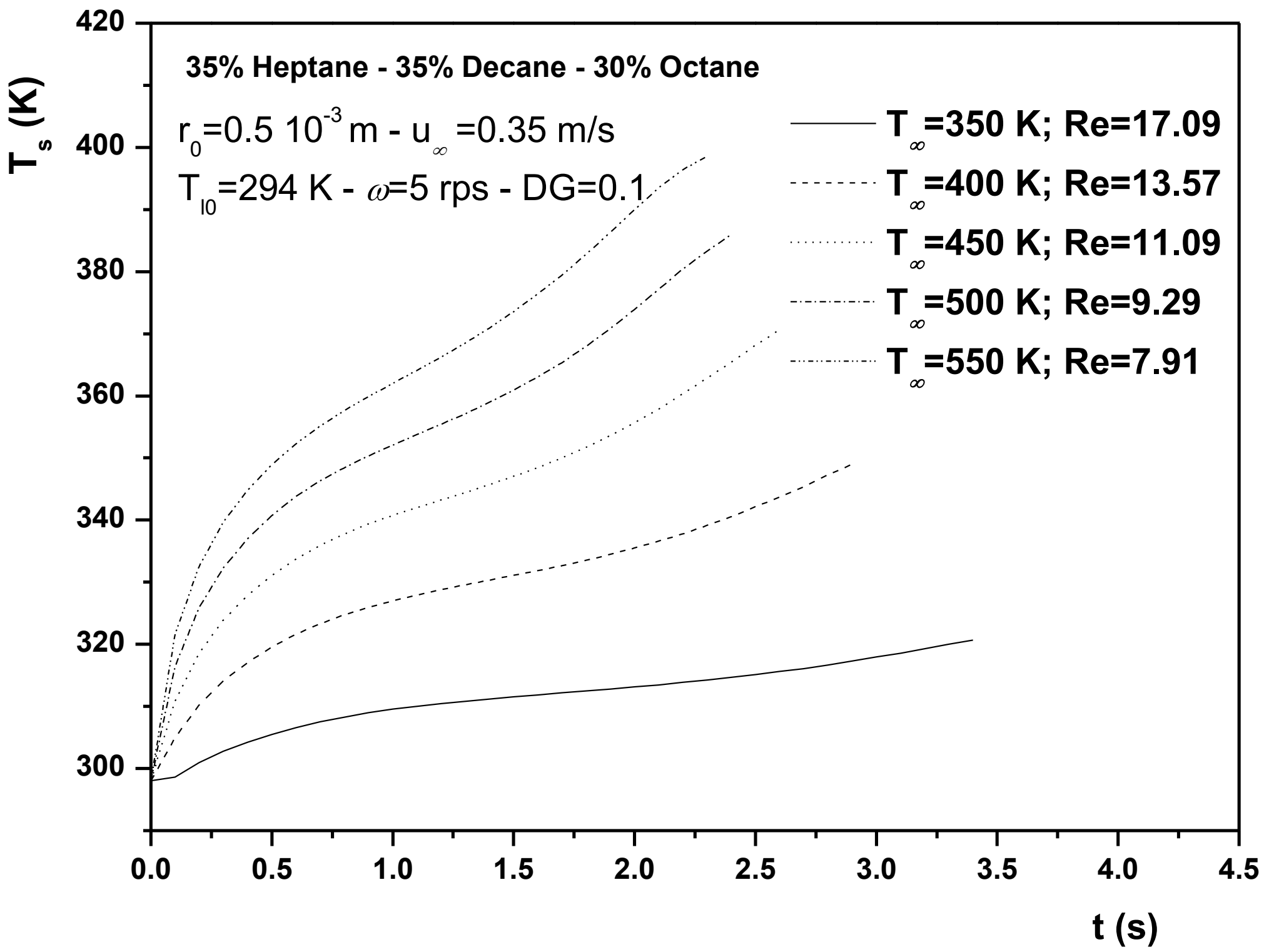

Figure 6: Time increase of the droplet surface temperature of ternary components (35% Heptane - 35% Decane – 30% Octane) in evaporation in rotatory forced convection for different ambient temperatures.

The effect of the evaporation in a hot wind of low velocity, and with a rotating droplet of 5 rps, accelerates the increase of the surface temperature of the liquid droplets. Therefore, the surface temperature of the liquid droplet reaches its boiling temperature. This phenomenon decreases rapidly the square diameter of the liquid droplets (*figure 7*). During this evolution, the $d^2$ law is confirmed for each component separately. However, the $d^2$ law is not confirmed for the three components mixed together. Furthermore, the regression of the square diameter of the liquid droplet follows three levels. The first level corresponds to the evaporation of the heptane component, followed by a second level corresponding to the evaporation of the octane component and the third one is due to the evaporation of the less volatile component (decane component). Nevertheless, the evaporation of the heptane mono-component liquid droplet, in a hot environment, accelerates the evaporation process without producing enough energy, because of the influence of the heptane’s low boiling temperature (371.58 K). In addition, the evaporation of the decane mono-component liquid droplet produces enough power due to its boiling temperature (447.3 K), but it slows the evaporation process. Finally, the evaporation of a rotating liquid hydrocarbon droplet of three components (heptane, octane and decane) produces enough power due to the impact of the decane's boiling temperature, and further, improves this phenomenon by accelerating the evaporation process due to the heptane liquid droplet evaporation. Thus, the preheating in an internal combustion engine of the fuels of three components improves the evaporation phenomenon of the rotating droplets.

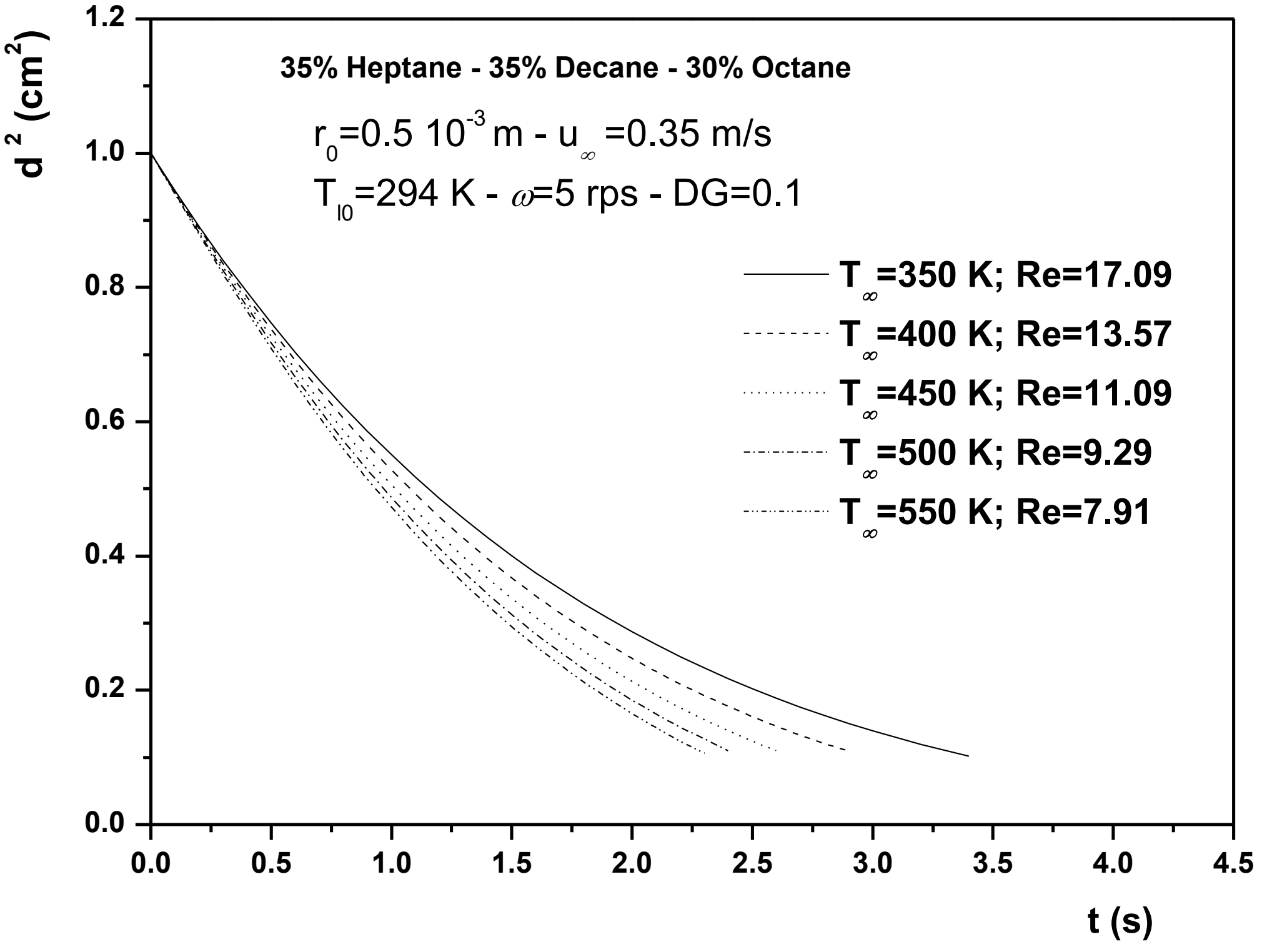

Figure 7: Time increase of the regression of the droplet square diameter of ternary components (35% Heptane - 35% Decane - 30% Octane) in evaporation in rotatory forced convection for different ambient temperatures.

Figures 8 and 9 show the influence of the wind velocity on the evaporation of a liquid hydrocarbon droplet of three components, in rotatory forced convection. These figures show that the rise of the wind velocity accelerates the evaporation phenomenon by reducing the time of the evaporation process. This phenomenon presents a quick rise of the surface temperature (*figure 8*), which remains below the boiling temperature of the less volatile component. In addition, the Reynolds number increases with the augmentation of the wind velocity while remaining in laminar regime.

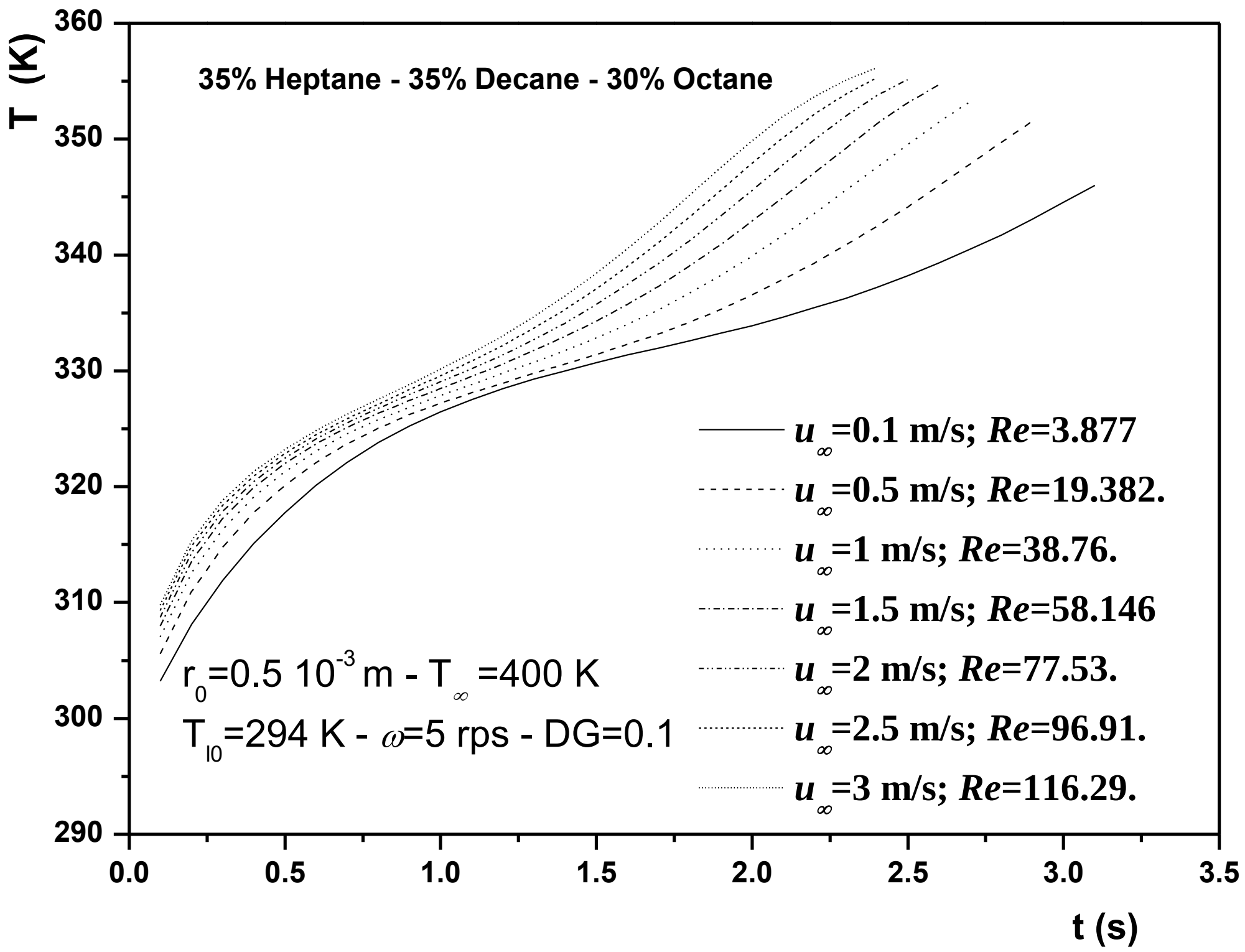

Figure 8: Time increase of the droplet surface temperature of ternary components (35% Heptane - 35% Decane - 30% Octane) in evaporation in rotatory forced convection for different air velocities.

Therefore, the figure 9 shows that the increase of the wind velocity evaporates rapidly the multi-component liquid droplet. These two figures also show the evaporation rate of each component and the evaporation rate of the hydrocarbons mixture. The evaporation rate of each component is determined by the slope that corresponds to the regression of the square diameter of the droplet as a function of time, as shown in figure 9.

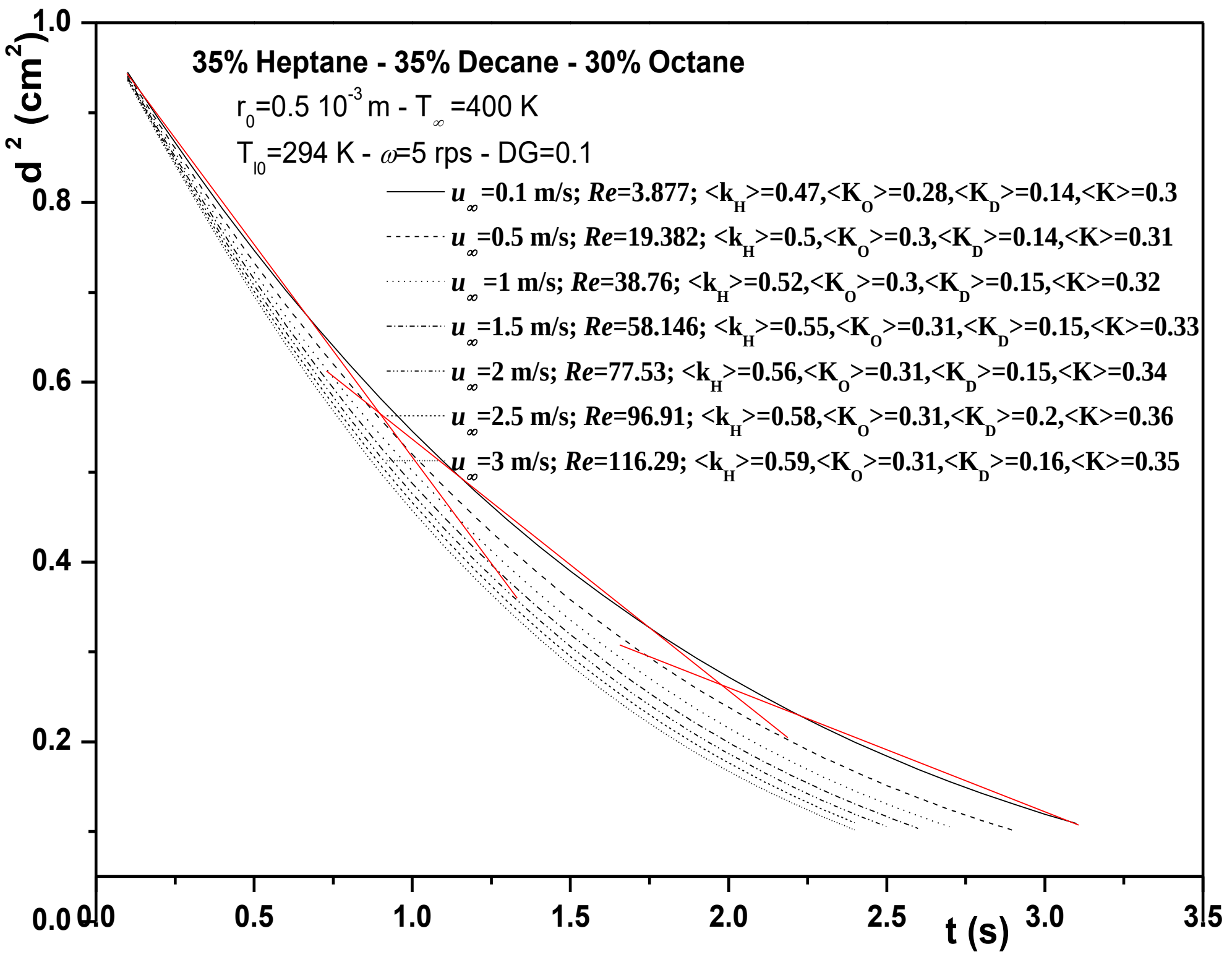

Figure 9: Time increase of the regression of the droplet square radius of ternary components (35% Heptane - 35% Decane - 30% Octane) in evaporation in rotatory forced convection for different air velocities.

Therefore, the evaporation of the liquid hydrocarbons droplet of three components contains three evaporation rates corresponding to three slopes. The average evaporation rate of the liquid hydrocarbon droplet of the three components is calculated as follows:

$$\bar{K} = 0.35 \times K_H + 0.30 \times K_O + 0.35 \times K_D \qquad (19)$$

Where $K_H$, $K_O$ and $K_D$ are the evaporation rates of the heptane, octane and decane respectively.

The average evaporation rate is calculated from the summation of the evaporation rate of each individual component, multiplied by the percentage of the initial mass fraction of the same component. The average evaporation rate increases with the augmentation of the wind velocity, and varies between 0.30 $mm^2$/s and 0.36 $mm^2$/s. This phenomenon indicates that the optimal value of the wind velocity (150 cm/s) accelerates the evaporation process.

In forced convection, the influence of the initial radius of a rotating liquid hydrocarbon droplet on the regression of the square diameter and the evolution of the surface temperature is presented in figures 10 and 11 respectively. The average evaporation rate that strongly depends on the variation of the initial radius, the wind velocity, and the rotational velocity of the liquid droplet, is determined. Figure 10 shows the evaporation rates for different initial radii. ($K_1$) is the evaporation rate of the first slope that corresponds to the evaporation of the heptane component, ($K_2$) is the evaporation rate of the second slope corresponding to the evaporation of the octane component, and ($K_3$) is the evaporation rate of the third slope corresponding to the less volatile component (decane). Thus, the average evaporation rate is determined as above for different initial radii. Figure 10 shows also that the average evaporation rate decreases with decreasing the droplet initial radius.

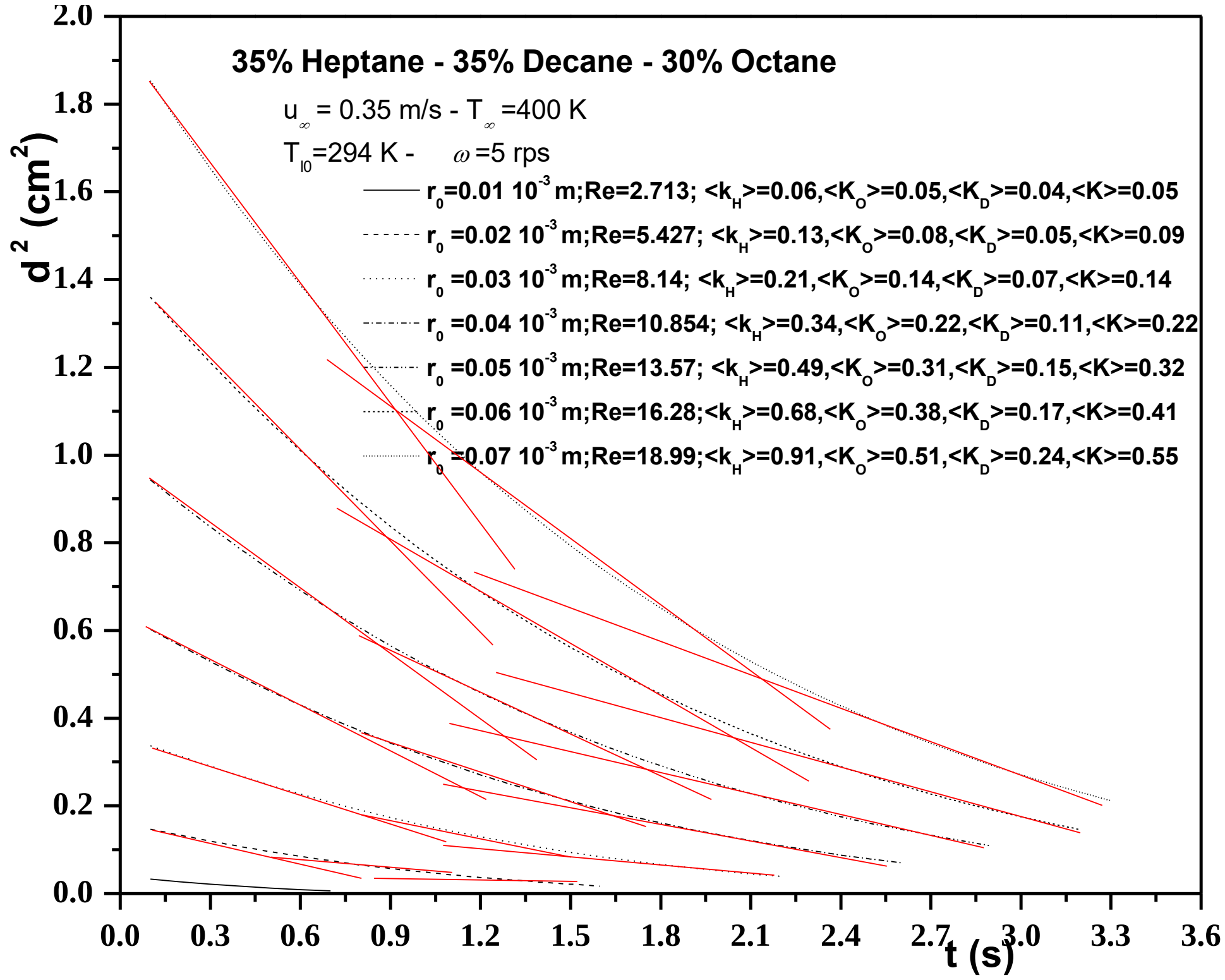

Figure 10: Time increase of the regression of the droplet square diameter of ternary components (35% Heptane - 35% Decane - 30% Octane) in evaporation in rotatory forced convection for different initial radii.

Figure 11 shows that the surface temperature of the rotating droplet of ternary components increases rapidly to reach the boiling temperature of the less volatile component (decane). However, the surface temperature needs more time to reach its boiling temperature for large droplets at a wind velocity of 0.35 m/s, an ambient temperature of 400 K, and a rotational velocity of 5 rps.

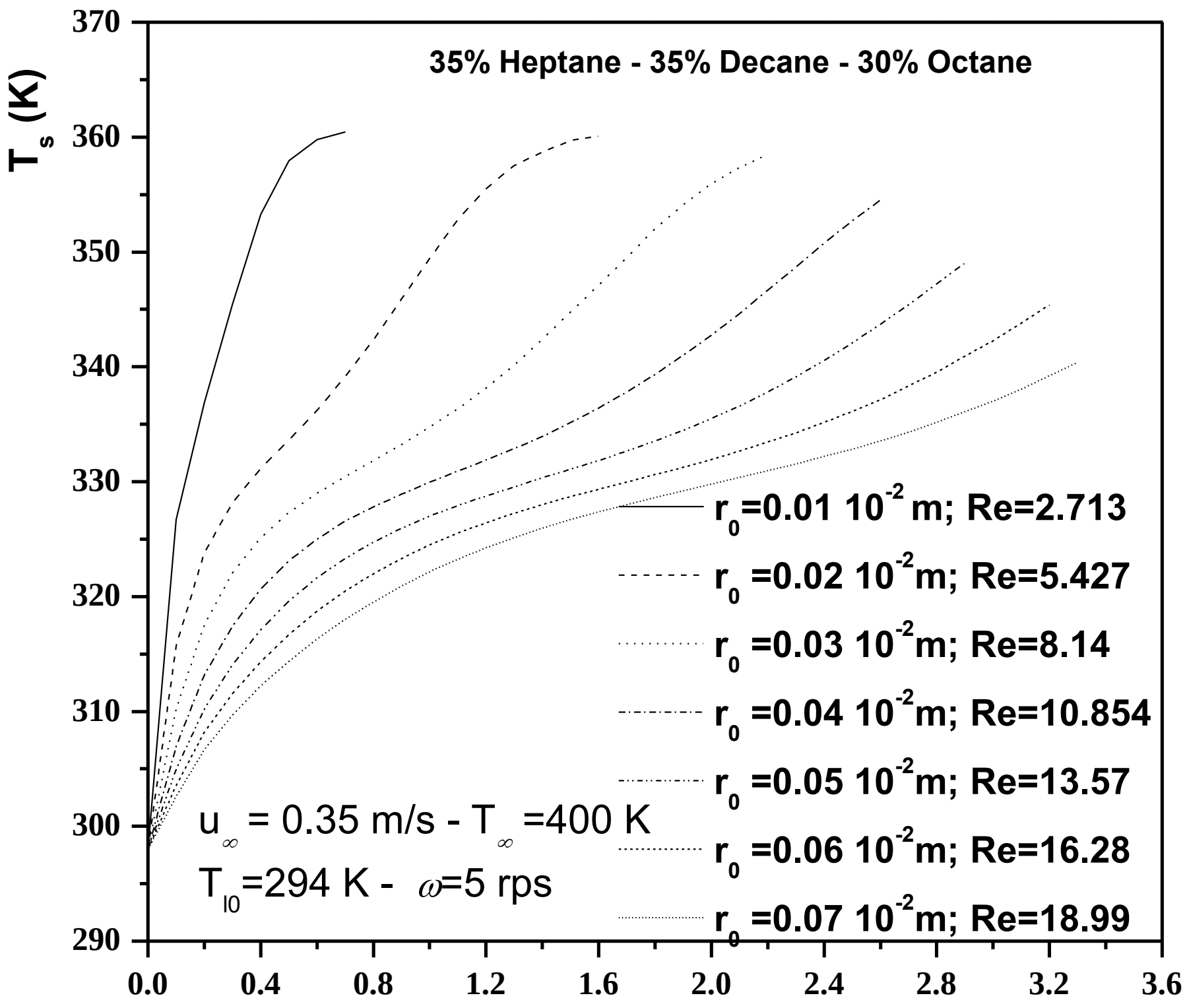

Figure 11: Time increase of the droplet surface temperature of ternary components (35% Heptane - 35% Decane – 30% Octane) in evaporation in rotatory forced convection for different initial radii.

Our objective is to determine new correlations for the average evaporation rate of the rotating hydrocarbon liquid droplet of ternary components. Firstly, the average evaporation rate for a hydrocarbon liquid droplet of multi-component blends without rotation, in forced convection, is determined. Thus, the figure 12 shows that the average evaporation rate for a stagnant liquid droplet increases with the increase of the air flow velocity varying from 0.048 $mm^2/s$ to 0.122 $mm^2/s$, by verifying the results of (Nomura et al, 1996) and (Chauveau et al, 2008). Furthermore, a new correlation expressing the average evaporation rate as a function of Reynolds number and the evaporation rate without convection is determined for a stagnant liquid droplet using the least squares method. This correlation is presented by the following equation:

$$K=0.0306+0.0093Re^{0.5} \quad (20)$$

It is also presented as follows:

$$(K-K_0)/K_0=0.3039Re^{0.5} \quad (21)$$

Where $K_0$=0.0306, is the evaporation rate, for the case without convection.

This correlation takes into consideration the variability of the thermo-physical and transport properties of a liquid hydrocarbon droplet of three components, the effect of the variation of the initial radius of the stagnant liquid droplet, and the effect of the variation of the air velocity in forced convection without rotation.

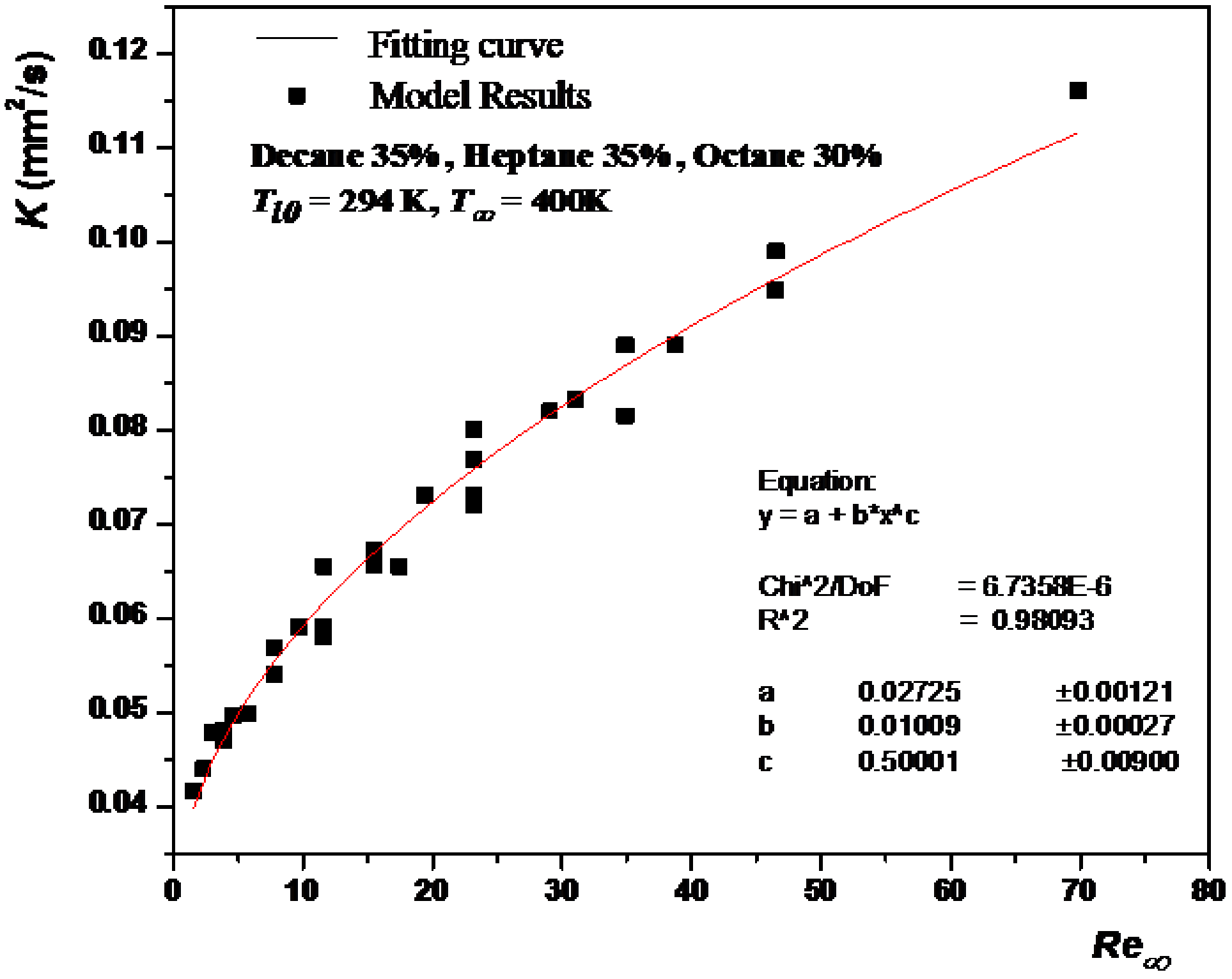

Figure 12: Average evaporation rate evolution versus Reynolds numbers in forced convection.

Secondly, by adding the rotation phenomenon to the evaporation of the liquid droplet of ternary components, the evaporation rate increases gradually with the increasing of the rotation velocity (*figure 13*). A new correlation expressing the average evaporation rate based on the Dgheim number is determined by using the least squares method as shown in figure 13. The resulting mathematical model is the following:

$$K = 0.02725 + 0.01009Re^{0.5} + 184.94048DG^{3} \tag{22}$$

This correlation takes into account the evaporation in forced convection of the rotating liquid hydrocarbon droplets of ternary components, wherein the thermo-physical and transport properties such as the thermal conductivity, density, specific heat, dynamic viscosity and diffusion coefficients, are variables. In addition, it takes into account the variation of the air velocity, and the size of the liquid droplets (initial droplet radius varying from 0.1 mm to 0.7 mm). For each droplet size, and each air velocity value, the rotation velocity varies from 1 rps to 30 rps. Thus, the average evaporation rate is calculated from the summation of the evaporation rate without convection, the evaporation rate in forced convection that depends on Reynolds number, and the evaporation rate in rotatory convection that depends on Dgheim number. It shows also that the rotatory convection is more predominant than the forced convection, and that the effect of the Dgheim number is higher than the effect of the Reynolds number on the evaporation of the rotating liquid hydrocarbon droplet. The average evaporation rate reaches values greater than 1 $mm^2/s$ for high value of Dgheim number, while the average evaporation rate reaches 0.12 $mm^2/s$ for high values of Reynolds number.

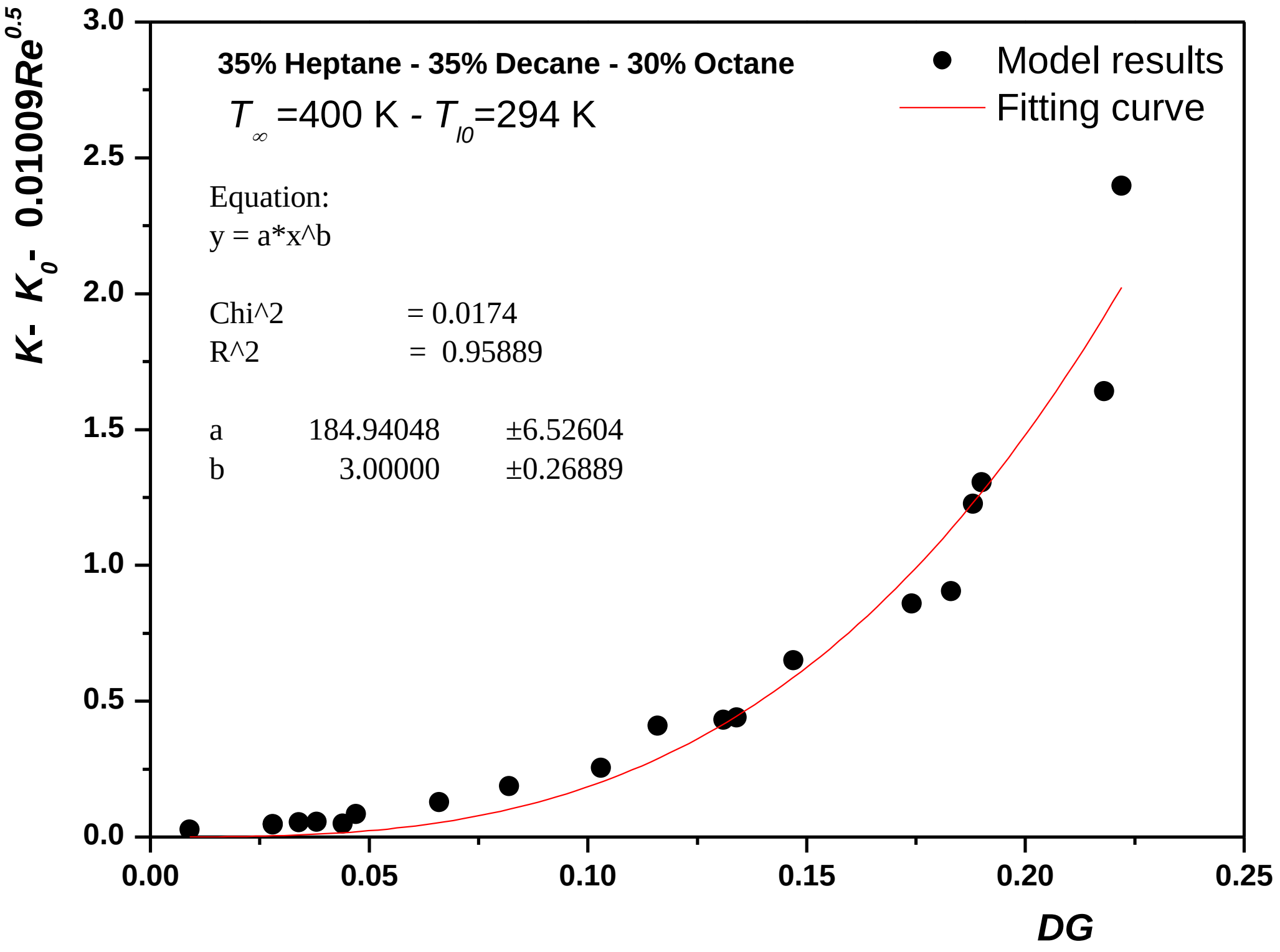

Figure 13: Average evaporation rate evolution versus Dgheim numbers in rotatory forced convection.

By performing the evaporation of a liquid hydrocarbon droplet of three components in rotatory convection and in an environment with different ambient temperatures, a new correlation expressing the average evaporation rate versus Dgheim and Prandtl numbers, is determined (*figure 14*). By using the least squares method, the new correlation is determined at a very small value of the airflow velocity. Thus, the effect of the forced convection (Reynolds number) is reduced. The mathematical form of this new correlation is the following:

$$K = 0.02725 + 70.26299(DG * Pr)^2 \quad (23)$$

Similarly, by performing the evaporation of a liquid hydrocarbon droplet of three components, in forced convection with different ambient temperatures, a new correlation expressing the evaporation rate versus Reynolds and Prandtl numbers, is determined (*figure 15*). By using the least squares method, the new correlation is determined at a very small value of the rotation velocity. Thus, the effect of the rotatory convection (Dgheim number) is reduced. The mathematical form of this new correlation is the following:

$$K = 0.02725 + 0.01542(Re * Pr)^{0.5} \quad (24)$$

The final correlation that takes into consideration the rotatory forced convection in an environment of different temperatures is written as following:

$$K = 0.02725 + 0.01542(Re * Pr)^{0.5} + 70.26299(DG * Pr)^2 \quad (25)$$

It can be written under this new form:

$$K = 0.02725 + 0.01542Pe_T^{0.5} + 70.26299Pe_{TR}^2 \quad (26)$$

Where, $Pe_T$ and $Pe_{TR}$ are respectively, the thermal Peclet number, and the thermal rotatory Peclet number.

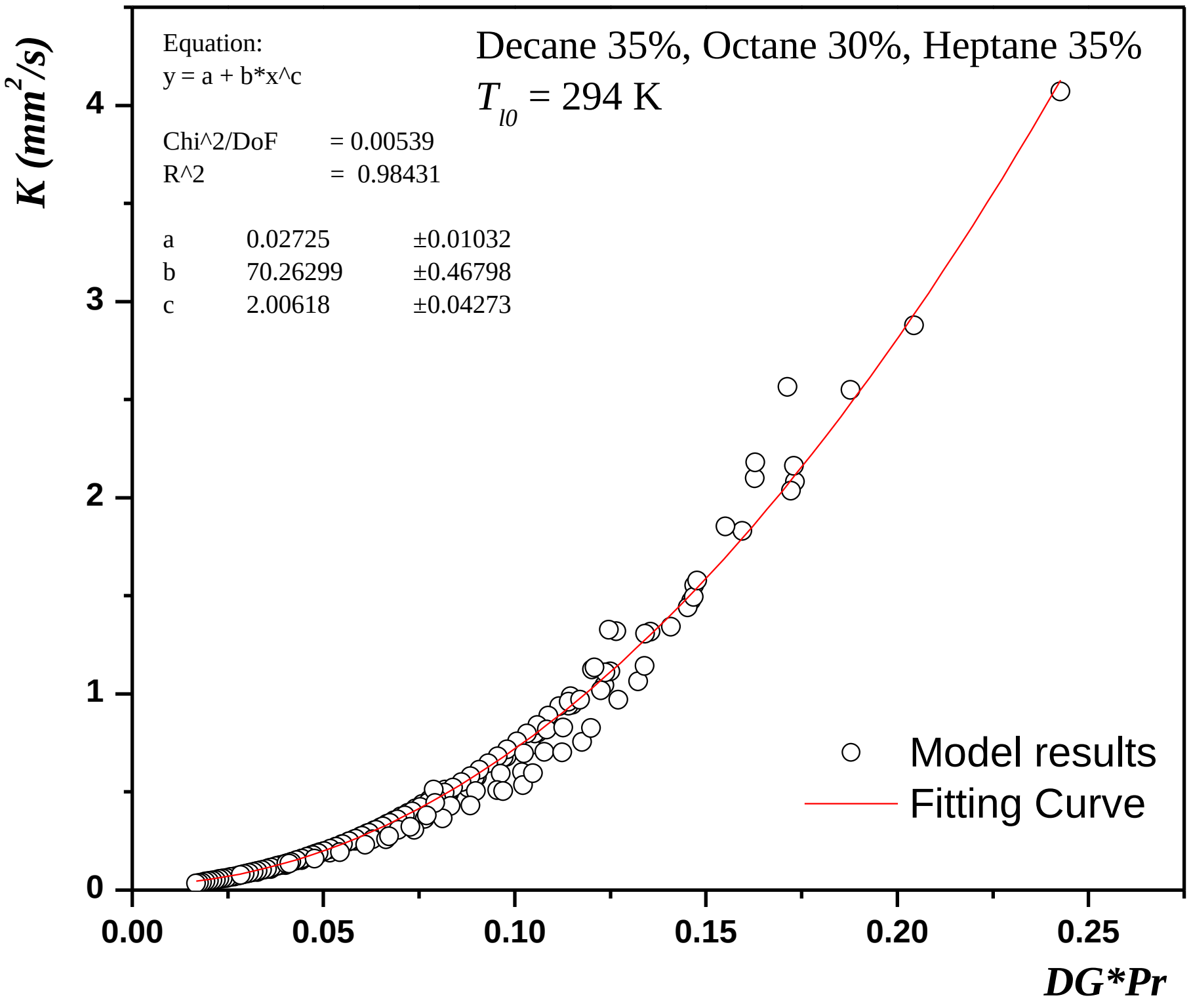

Figure 14: Average evaporation rate evolution versus Dgheim and Prandtl numbers in rotatory convection.

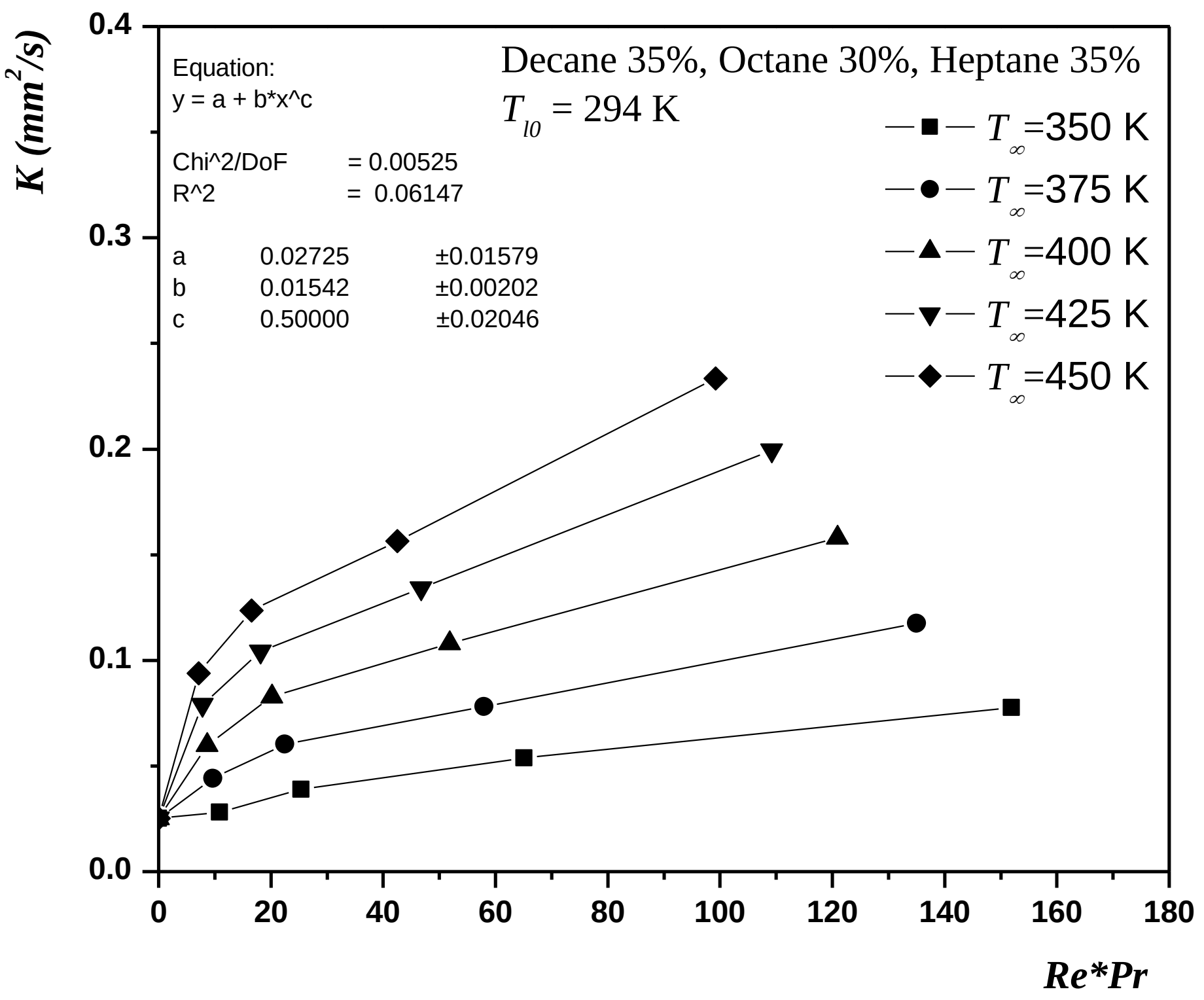

Figure 15: Average evaporation rate evolution versus Reynolds and Prandtl numbers in forced convection.

Thus, the average evaporation rate for different ambient temperatures is calculated from the summation of the evaporation rate without convection, the evaporation rate in forced convection and the evaporation rate in rotatory convection.

Moreover, by performing the evaporation of a liquid hydrocarbon droplet of three components in rotatory convection, with different initial mass fractions, a new correlation expressing the average evaporation rate versus Dgheim and Schmidt numbers, is determined (*figure 16*). By using the least squares method, the new correlation is determined at a very small value of the airflow velocity. Thus, the effect of the forced convection (Reynolds number) is reduced. The mathematical form of this correlation is the following:

$$K = 0.02725 + 2.18544(DG * Sc)^{1.5} \quad (27)$$

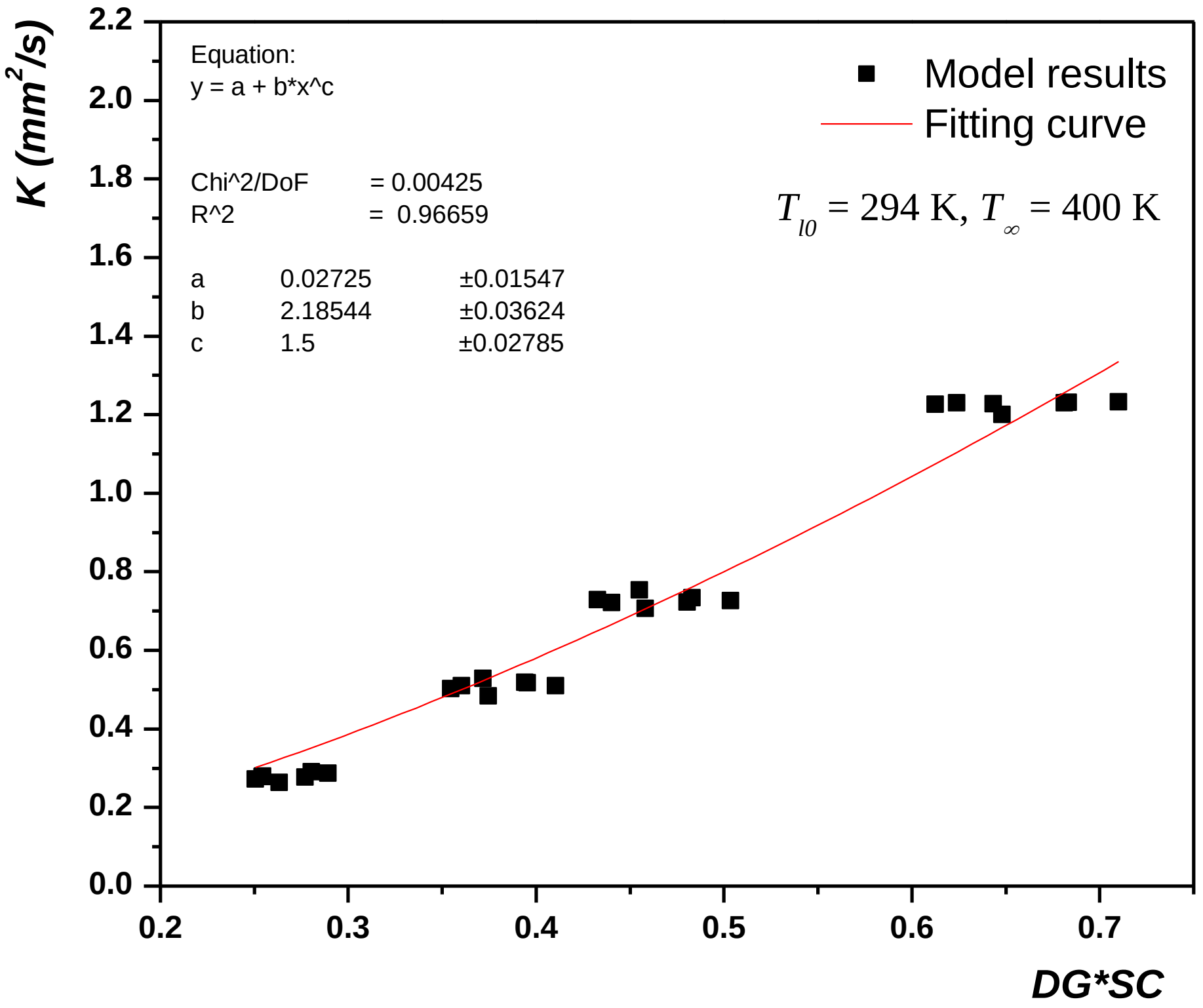

Figure 16: Average evaporation rate evolution versus Dgheim and Schmitt numbers in rotatory convection.

Similarly, by performing the evaporation of a liquid hydrocarbon droplet of three components in forced convection, with different initial mass fractions, a new correlation expressing the average evaporation rate versus Reynolds and Schmidt numbers, is determined (*figure 17*). By using the least squares method, the new correlation is determined at a very small value of the rotation velocity. Thus, the effect of the rotatory convection (Dgheim number) is reduced. The mathematical form of this correlation is the following:

$$K = 0.02725 + 0.00689(Re * Sc)^{0.5} \quad (28)$$

The final correlation that takes into consideration the rotatory forced convection is written as following:

$$K = 0.02725 + 0.00689(Re * Sc)^{0.5} + 2.18544(DG * Sc)^{1.5} \quad (29)$$

It can be written under this new form:

$$K = 0.02725 + 0.01542Pe_M^{0.5} + 70.26299Pe_{MR}^{1.5} \quad (30)$$

Where, $Pe_M$ and $Pe_{MR}$ are respectively, the mass Peclet number, and the mass rotatory Peclet number.

Thus, the average evaporation rate for different initial mass fractions is calculated from the summation of the evaporation rate without convection, the evaporation rate in forced convection and the evaporation rate in rotatory convection.

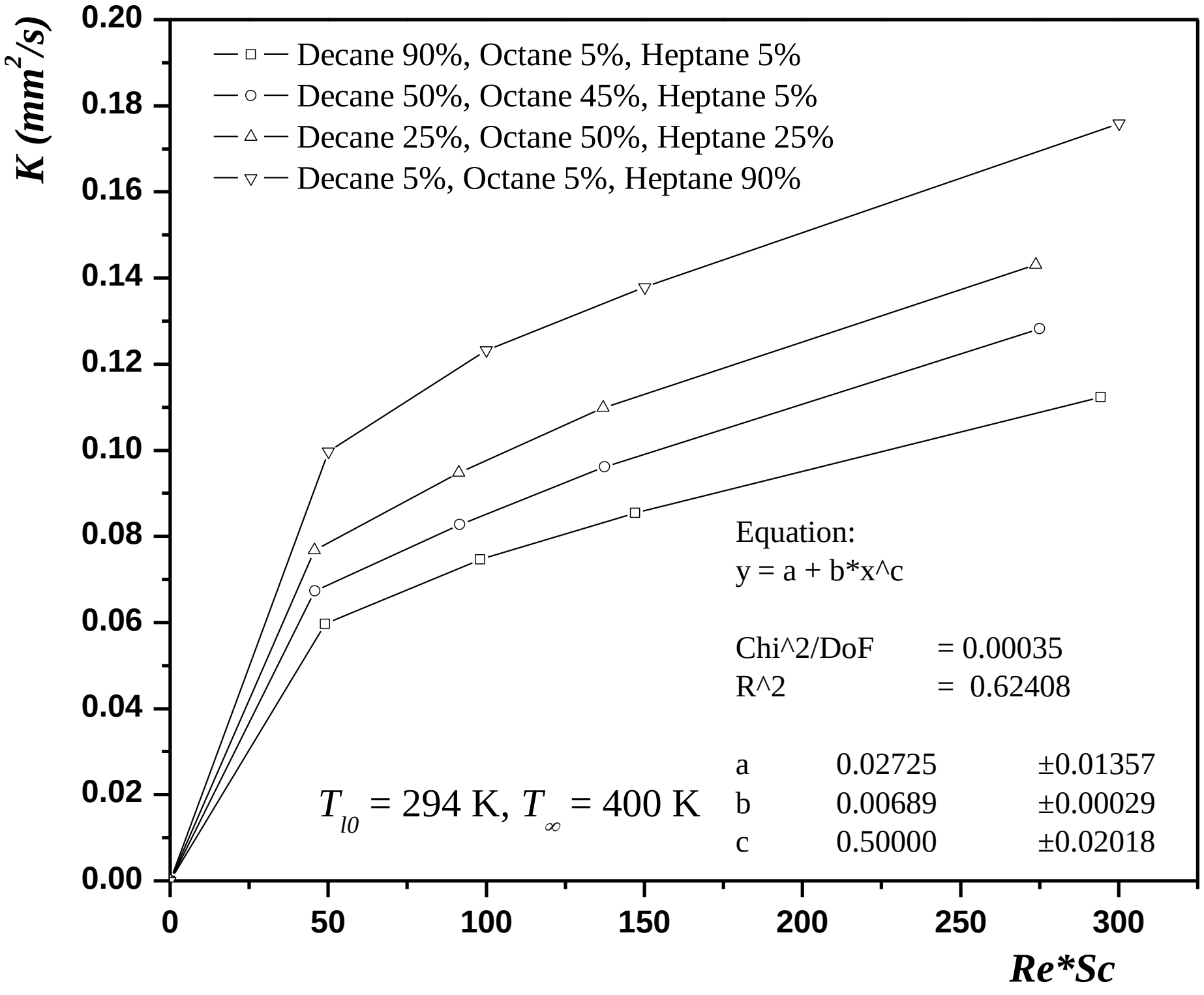

Figure 17: Average evaporation rate evolution versus Reynolds and Schmitt numbers in forced convection.

## 4. Conclusion

A numerical model describing the heat and mass transfers of a hydrocarbon liquid droplet of three components in rotatory forced convection is studied. This model treats the evaporation of a hydrocarbon liquid droplet in rotation under the effect of a hot air flow. The mathematical model consists on a system of algebraic equations. The numerical model uses the implicit finite difference method and Thomas algorithm. The numerical model treats the heat and mass transfers' equations by taking into account the variability of the thermo-physical and transport properties of a hydrocarbon liquid droplet of ternary components.

The evaporation of a hydrocarbon liquid droplet of ternary components shows phenomena that do not appear in the case of the evaporation of a liquid droplet of one or two components. The obtained results are as follows:

- In this model, the evaporation of the rotating liquid droplets in forced convection is studied. Then, Nusselt and Sherwood numbers are determined to be a function of three terms: a first term is related to the evaporation of a stagnant spherical droplet equal to 2, a second term is related to the evaporation of the liquid droplet in forced convection as proposed by Renksizbulut and Yuen, and a third term is related to the evaporation of the liquid droplet under the effect of the rotation as proposed by Dgheim et al.

- A good qualitative and quantitative agreements are observed between our results for two components and those obtained by Bouaziz et al for the regression of the square diameter and the evolution of the surface temperature of the liquid droplet. In addition, the numerical model of three components (with and without rotation) shows a satisfactory agreement with that achieved for the above mentioned results.

- The profile of the square diameter of the liquid droplet, of three components, for a low value of Reynolds number, does not follow the $d^2$ law, and has three evaporation rates. The average evaporation rate decreases with the decreasing of the value of the droplet initial radius and increases with the increasing of the wind velocity, and the rotation velocity of the liquid droplet.

- An optimal value of the wind velocity (1.5 m/s), in rotatory forced convection, accelerates and improves the process of the evaporation.

- The evaporation in a hot environment accelerates the increase of the surface temperature of the rotating liquid droplets and, therefore, reaches rapidly its boiling temperature. Moreover, it reduces quickly the droplet radius. Therefore, the pre-heating in an internal combustion engine improves the evaporation phenomenon of the droplets.

- An Optimal value of the rotational velocity (15 rps) of the liquid droplet improves the evaporation phenomenon by increasing its evaporation rate.

- New correlations expressing the evaporation rate in terms of Reynolds, Dgheim, Schmitt and Prandtl numbers, and the initial evaporation rate (without convection) are determined. These correlations take into account the variability of the thermo-physical and transport properties of the liquid hydrocarbon droplets of ternary components, in rotatory forced convection. They also take into consideration the variation of the wind velocity, and the spinning frequency of the liquid droplet of variable initial radii, and initial mass fractions.

- The average evaporation rate for heat and mass transfers is calculated from the summation of the evaporation rate without convection, the evaporation rate in forced convection (Peclet number) and the evaporation rate in rotatory convection (rotatory Peclet number).
- Finally, the influence of the Reynolds number on the average evaporation rate of a hydrocarbon liquid droplet of ternary components, in rotation, is almost minimal, but it is not negligible comparing to the effect of the Dgheim number.

**Acknowledgement:**

*This work was supported by the Lebanese National Scientific Research Center CNRS.*

## Nomenclature

### *Symbols*

| | | |
|---|---|---|
| *B* | Spalding transfer number | |
| *Cp* | specific heat | $J.kg^{-1}.K^{-1}$ |
| *D* | diffusion coefficient | $m^2.s^{-1}$ |
| *DG* | Dgheim number | |
| *f* | spinning frequency | $s^{-1}$ |
| *K* | evaporation rate | $mm^2.s^{-1}$ |
| $<K>$ | average evaporation rate | $mm^2.s^{-1}$ |
| *L* | latent heat | $J.kg^{-1}$ |
| *M* | molar mass | $kg.kmol^{-1}$ |
| *Nu* | Nusselt number | |
| *P* | Pressure | atm |
| *Pe* | Peclet number | |
| *Pr* | Prandtl number | |
| *r* | sphere radius | m |
| *Re* | *Reynolds number* | |
| *Sc* | Schmidt number | |
| *Sh* | Sherwood number | |
| *t* | time | s |
| *T* | temperature | K |
| *u* | velocity component in x direction | $m.s^{-1}$ |
| *Y* | mass fraction | |

Greek letters

| | | |
|---|---|---|
| $\eta$ | $r/r_s$ | |
| $\lambda$ | thermal conductivity | $W.m^{-1}.K^{-1}$ |
| $\nu$ | Kinematic viscosity | $m^2.s^{-1}$ |
| $\rho$ | density | $kg.m^{-3}$ |
| $\omega$ | angular velocity | rps |

Subscripts

| | |
|---|---|
| $a$ | air |
| $crit$ | critical |
| $f$ | fuel |
| $ebn$ | boiling |
| $fT$ | heat transfer limit |
| $fM$ | mass transfer limit |
| $g$ | gas |
| $i$ | component |
| $j$ | component |
| $L$ | liquid |
| $l$ | liquid |
| $M$ | mass |
| $R$ | rotatory |
| $s$ | surface |
| $T$ | thermal |
| $VS$ | saturated vapor |
| $V$ | vapor |
| $\infty$ | ambient medium |
| $0$ | initial |

Superscripts

| | |
|---|---|
| $\bar{}$ | mean |
| * | defined |